\documentclass[12pt]{iopart}
\usepackage{epsfig}
\usepackage{amssymb}
\usepackage{bm}
\begin{document}

\title[Contractile tumor nodules]{Myosin-II dependent cell contractility contributes to spontaneous nodule formation of mesothelioma cells. }

\author{ Julia T\'arnoki-Z\'ach$^1$,
	 Dona Greta Isai $^{2}$,
	 El\H od M\'ehes$^1$,
	 S\'andor Paku,
	 Zolt\'an Neufeld,
	 Bal\'azs Heged\H us,
	 Bal\'azs D\"ome,
 Andras Czirok $^{1,2}$\footnote{corresponding author} ,
       }
\address{$^1$ Department of Biological Physics, Eotvos University, Budapest, Hungary}
\address{$^2$ Department of Anatomy \& Cell Biology, University of Kansas Medical Center, Kansas City, KS, USA}

\ead{aczirok@gmail.com}

\begin{abstract}

We demonstrate that characteristic nodules emerge in cultures of several  malignant pleural mesothelioma (MPM) cell lines. Instead of excessive local cell proliferation, MPM nodules arise in culture by Myosin II-driven cell contractility. Accordingly, the aggregation process can be prevented or reversed by suitable pharmacological inhibitors of Myosin II activity. A cell-resolved elasto-plastic model of the multicellular patterning process predicts that the morphology and size of the nodules as well as the speed of their formation is determined by the mechanical tension cells exert on their neighbors, and the stability of cell-substrate adhesion complexes.  For small tension forces nodules are slow to develop and localized at the boundary of cell free areas. In contrast, a high intralayer tension quickly transforms all cell-covered areas into dense clusters interconnected by multicellular strands. Model simulations also indicate that a decreased stability of cell-substrate adhesions favors the  formation of fewer, but larger clusters.  Linear stability analysis of a homogenous, self-tensioned Maxwell fluid indicates the unconditional presence of the patterning instability.

\end{abstract}

\vspace{2pc}

\maketitle

\def\be{\begin{equation}}
\def\ee{\end{equation}}
\def\bea{\begin{eqnarray}}
\def\eea{\end{eqnarray}}
\def\r{\mathbf{r}}
\def\p{\mathbf{p}}
\def\hd{\hat{\mathbf{d}}}
\def\x{\mathbf{x}}
\def\n{\mathbf{n}}
\def\t{\mathbf{t}}
\def\F{\mathbf{F}}
\def\tF{\tilde{F}}
\def\tE{\tilde{E}}
\def\tC{\tilde{C}}
\def\tk{\tilde{k}}
\def\td{\tilde{d}}
\def\talpha{\tilde{\alpha}}
\def\P{\mathbf{P}}
\def\Q{\mathbf{Q}}
\def\M{\mathbf{M}}
\def\N{\mathbf{N}}
\def\R{\mathbf{R}}
\def\V{\mathbf{V}}
\def\v{\mathbf{v}}
\def\u{\mathbf{u}}
\def\bphi{\bm{\phi}}

\section{Introduction}
Cell contractility and forces exerted on the cell's microenvironment constitute an important mechanism of multicellular patterning. Morphogenetic movements in epithelia are particularly well known to utilize cell contractility: for example, an anisotropic contractile activity gives rise to cell intercalation, a process altering cell connectivity/adjacency in such a way that the whole tissue elongates in one direction while narrows along the perpendicular direction \cite{Bertet04,Honda08}. Similarly, contractility-driven constriction of the free (apical) surface of the epithelium can give rise to bending or budding within an epithelial sheet \cite{Martin14}. Mesenchymal cells can also contract the surrounding extracellular matrix (ECM): if a cell aggregate is placed on the surface of a collagen gel, cell traction reorganizes the collagen and create bundles of ECM that radiate away from the aggregate \cite{SH82, Sawhney02}. This observation led to the development of the mechano-chemical theory of pattern formation \cite{Murray83,Oster83}, according to which cells exert traction forces on an underlying deformable substrate and the resulting strain transports (convects) both cells and the ECM. Furthermore,  strain-oriented ECM filaments can guide cell motility, as cells are more likely to move parallel with the orientation of the ECM \cite{Murray98, Murray2, Manoussaki96}. This mechanism was suggested to guide vascular patterning, and endothelial cells were reported to be able to detect and respond to substrate strains created by the traction stresses of neighboring cells \cite{Reinhart-King08}. 

Cellular contractility is also an important factor in tumor progression as it can contribute both to local and distant spreading of malignant cells \cite{Bhandary15, Mierke11, Poincloux11, Krndija10}. Here we focus on the role of cell contractility in malignant pleural mesothelioma (MPM), a tumor arising from mesothelial cells lining the pleural cavity. This highly aggressive disease has an incidence of 1 in 100 000 in Europe \cite{Lin07}. Despite recent advances in the treatment of MPM, it has an extremely poor prognosis with almost all patients dying from their tumor. A characteristic, pathognomonic feature of MPM is the formation of multiple, macroscopic pleural tumor nodules, which -- due to the special two dimensional environment -- may pinch off and contribute to the local spreading of malignant cells. 

In this manuscript we demonstrate that several MPM cell lines can form nodule-like aggregates in vitro when cultured at high cell density. Time-lapse analysis of their formation and experiments with myosin II-specific inhibitors reveal that cell contractility is a key factor in the nodule formation process. To interpret these experimental findings, we propose a cell-resolved elasto-plastic model of a contractile cell layer. By computational simulations we explore the effects of key model parameters, such as the the stability of cell-cell and cell-substrate adhesion and the magnitude of the cell-exerted contractile forces.

\section{Methods}
\subsection{Cell lines}
SPC111 cells were established from human biphasic MPMs and kindly provided by Prof. R. Stahel (University of Zurich, Switzerland). P31 cells were a kind gift from Prof. K. Grankvist (University of Umea, Sweden). The Meso cell lines were established by the Vienna MPM group as described recently \cite{Garay13}. NP3 normal mesothelial cells were isolated from pneumothorax patients using the same protocol \cite{Laszlo15}.

\subsection{Culture conditions}
Gelatin-coated dishes were obtained by incubating 10\% gelatin solution-B (Sigma) in PBS for 45 min at room temperature. We also prepared 1.7 mg/ml Collagen-I (Corning) gels according to the manufacturer's instructions. 

Cells were grown at 37$^o$C in a humidified, 5\% CO$_2$, 95\% air atmosphere. The DMEM (Lonza) medium was supplemented with 10\% FCS (Invitrogen) and 1\% penicillin-streptomycin-amphotericin B (Lonza). MPM cells were cultured on plastic tissue culture substrates (Greiner).  NP3 cells are, however, more sensitive to environmental factors and have a limited proliferative potential. Thus, NP3 cells were cultured in gelatin-coated dishes, for up to three passages.

\subsection{Reagents}
To interfere with normal Myosin II function, we utilized Y27632, the rho kinase (ROCK) inhibitor (Merck Millipore) and Blebbistatin (Merck Millipore), an inhibitor of actomyosin crosslinking. Y27632 was solved in destilled water, stored in the form of a 10 mM stock solution, and used at 100 $\mu$M final concentrations in DMEM. Blebbistatin was solved in DMSO, stored as 50 mM aliquots, and used at 40-80 $\mu$M final concentrations.

\subsection{Immunostaining}
SPC111 cells were seeded at confluent density (40000 cells/cm$^2$) either on the surface of collagen gels or on glass coverslips and were grown for 6 days. Samples were fixed using 4\% PFA for 15 min at 4$^o$C, permeabilized with 0.25\% Triton-X100 for 10 min at 4$^o$C, and incubated with the following antibodies: Polyclonal anti-fibronectin (1:100, Ab2033, Millipore); polyclonal anti-beta-catenin (1:100, C2206, Sigma). After washing, sections were incubated for 30 min with anti-rabbit Alexa-488 secondary antibody (Life Technologies, Carlsbad, CA). Filamentous actin was stained by Phalloidin-TRITC (P1951, Sigma). Samples were analyzed by confocal laser scanning microscopy using the Bio-Rad MRC-1024 system (Bio-Rad, Richmond, CA). 

\subsection{Physical sections}
Physical cross sections were obtained from cultures grown on the surface of collagen-I gels.  Samples were embedded in Spurr's mixture. Semithin (1 $\mu$m) sections were cut perpendicularly to the surface of the collagen and  stained with 0.5\% Toluidin blue. Images were taken using Zeiss Axioskop 2 microscope equipped with a 100x objective and coupled to an Olympus DP50 camera.

\subsection{Time-lapse microscopy}
Time-lapse recordings were performed on a computer-controlled Leica DM IRB inverted microscope equipped with a Marzhauser SCAN-IM powered stage and a 10x N-PLAN objective with 0.25 numerical aperture and 17.6 mm working distance. The microscope was coupled to an Olympus DP70 color CCD camera. Cell cultures were kept at 37°C in humidified 5\% CO$_2$ atmosphere during imaging. Phase contrast images of cells were collected consecutively every 10 minutes from each of the microscopic fields.

To maintain high cell density cultures for several days, we 3D printed three 6 mm diameter mini-wells into regular 35 mm tissue culture dishes. The side wall of the wells was formed by fused polylactic acid filaments. This setup allowed us to achive high (10$^6$/cm$^2$) cell density with a medium/cell ratio characteristic for culture with ten times lower cell density (10$^5$/cm$^2$). The 3 ml medium was replaced every 5 days. Four modified culture dishes, each containing 3 wells, were observed by time-lapse videomicroscopy for up to 14 days. We recorded four adjacent fields from each well, thus we collected 48 images in each imaging cycle (every 10 minutes).

\subsection{Computational model}
\begin{figure}
\begin{center}
\includegraphics[width=5in]{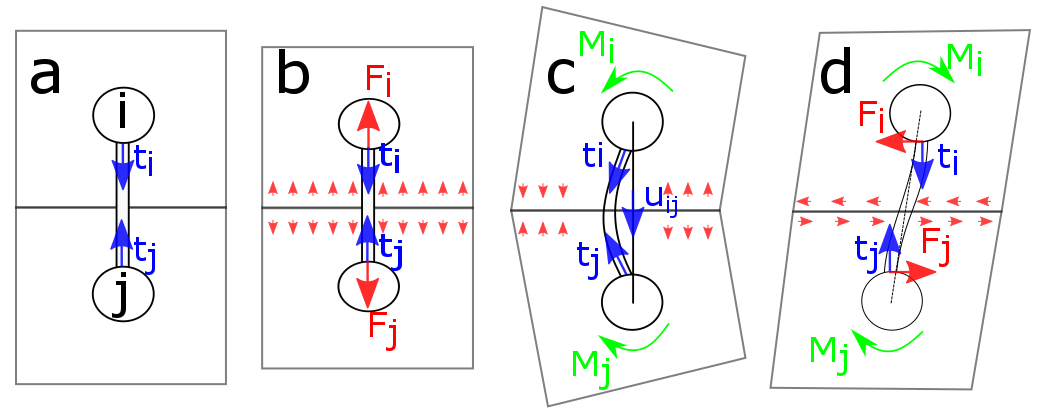}
\caption{\small
Mechanical model of multicellular clusters. Links represent mechanically connected cytoskeletal structures of adherent cells. Rectangular shapes indicate cell membranes, not explicitly resolved in the model.
a: Two particles, $i$ and $j$ interconnected with a link in a mechanical stress-free configuration. Unit vectors, $\t_i$ and $\t_j$, which co-rotate with the particles, represent the preferred direction of the link. b: Compressed cells. The interaction of the two cells gives rise to spatially distributed forces (red arrows), which are replaced by the repulsive net forces $\F_i$ and $\F_j$. c: A symmetric rotation of both particles yields torques $\M_i$ and $\M_j$ acting on particles $i$ and $j$, respectively. These torque vectors are perpendicular to the plane of the figure. d: A lateral misalignment of the particles creates torques $\M_i$, $\M_j$ and also  shear forces $\F_i$, $\F_j$ acting at the particles.  
}
\label{fig_model} 
\end{center}
\end{figure}

To explore the dynamics of cell contraction-driven aggregation and nodule formation, we adapted our cell-resolved elasto-plastic computational framework \cite{Czirok14} into a two dimensional model. Briefly, forces distributed along the membranes of mechanically coupled adherent cells are replaced by a single net force and a torque, simplifying the system to a network of particles and elastic beams (Fig.~\ref{fig_model}). 

In the computational model torques and shear forces arise due to relative movement of adjacent particles. Hook's law determines the force $F^\parallel$ needed to uniaxially compress or stretch cells (Fig.~\ref{fig_model}b). In particular, $F_l^\parallel$ is associated with changing the length of link $l$ which connects particles $i$ and $j$ located at $\r_i$ and $\r_j$, respectively:
 \be 
 F_l^\parallel = k(|\r_j-\r_i|-\ell_l).
 \label{Hook}
 \ee
In Eq. (\ref{Hook}) $\ell_l$ is the equilibrium length of link $l$ and $k>0$ is a model parameter, characterizing the stiffness of the cytoskelon. 

A link $l$ exerts a torque if its preferred direction at particle $i$, $\t_{i,l}$, is distinct from its actual end-to-end direction $\u_{ij}$ (Figs.~\ref{fig_model}c,d). We assume that the torque exerted on particle $i$ is proportional to the difference between the preferred and actual directions:
 \be
 \M_{i,l}=g (\t_{i,l} \times \u_{ij} ),
 \label{Mbend}
 \ee
where the microscopic bending rigidity $g$, a model parameter, can be calibrated from macroscopic material properties of the tissue \cite{Czirok14}. The condition for mechanical equilibrium, together with Eqs (\ref{Hook}) and (\ref{Mbend}) allow the calculation of the forces and torques within the system \cite{Czirok14}.

Multicellular plasticity is modeled using rules that rearrange the network of intercellular adhesions. The stability of intercellular adhesion complexes depends on the tensile forces they transmit \cite{Zhang04}. Thus, in our model the probability of removing link $l$ during a short time interval $\Delta t$ follows Bell's rule \cite{Bell78} as 
 \be
 p_l\Delta t = A e^{F_l^t/F^0} \Delta t,
 \label{link_removal}
 \ee 
where $F^0$ is a threshold value, and
 \be
 F_l^t=\Theta(F_l^\parallel)F_l^\parallel
 \label{fl}
 \ee
is the tensile component of the force transmitted by the link and $A$ is a scaling factor which characterizes the stability of connections. 

Mechanical connections can be established between two Voronoi neighbor particles, $i$ and $j$. We assume, that during a short time interval $\Delta t$ the probability of inserting a new link is a decreasing function of the distance $d_{i,j}$ between the particles:
 \be
 q_{i,j}\Delta t = B \left(1 - {d_{i,j} \over d_{max} } \right) \Delta t.
 \label{link_insert}
 \ee
The scaling factor $B$ represents the intensity of cellular protrusive activity devoted to scanning the environment and the ability to form intercellular contacts. The maximal distance cells explore for new connections is denoted by $d_{max}$. 

Simulations are event-driven: using the probability distributions (\ref{link_removal}) and (\ref{link_insert}), we generate the next event $\mu$ and waiting time $\tau$ according to the stochastic Gillespie algorithm \cite{Gillespie77}. The waiting time until the next event is chosen from the distribution
 \be
 \log P(\tau) = -\tau\left( \sum_l p_l + \sum_{i,j} q_{i,j} \right),
 \ee
where the sums are evaluated by iterating over existing links $l$ as well as over all possible Voronoi neighbor particle pairs $i,j$ not connected by a link.

\subsection{Simulation Parameters}
Two dimensional initial conditions were generated by randomly positioning $N=400$ particles in a square of size $L=20$. The unit distance of the simulations was set to the average cell size, $d_0\approx$10$\mu$m, thus the 2D cell density is 1 cell/unit area. In the initial condition we enforced that the distance of two adjacent particles is greater than $d_{min}=0.8d_0$. Particles that are Voronoi neighbors are connected by links when their distance is less than $d_{max}=2d_0$. For a mechanical stress-free initial configuration we set the $\t_{i,l}$ preferred link direction vectors as well as the equilibrium link lengths $\ell_l$ so that no internal forces or torques are exerted in the system. As a boundary condition, we fixed the position and orientation of the particles located near the perimeter of the simulation domain.

The mean waiting time between simulation events is set by parameters $A$ and $B$. We choose our time unit as $1/B\approx 10$ min, the time needed for two adjacent cells to establish a mechanical link.  We set the lifetime of an unloaded link to  $1/B \approx 1$ day. Thus, according to these values, two cells pulled away by a force $F^0$ separate in $\sim10$ h, a characteristic value consistent with the time scale observed in our cell culture experiments.

\section{Results}
\subsection{MPM cells form nodules in vitro}
\begin{figure}
\begin{center}
\includegraphics[width=5in]{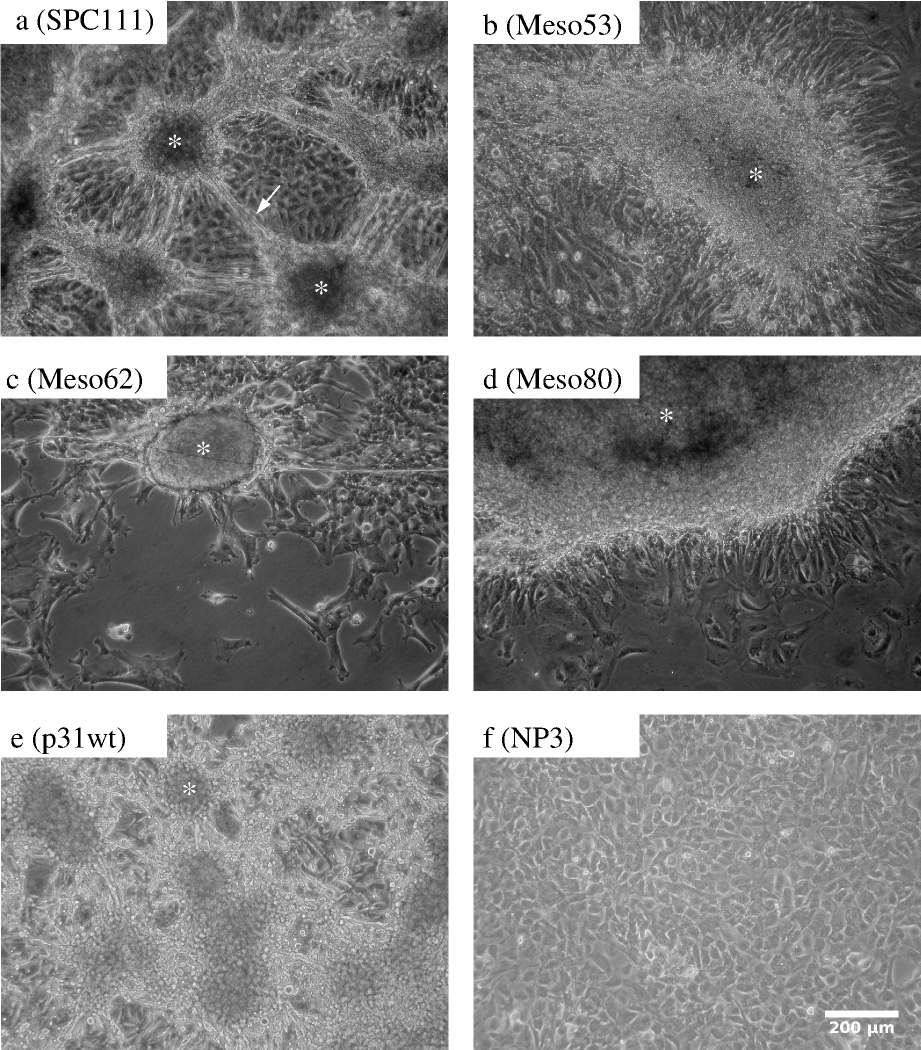}
\caption{\small
Mesothelioma cells spontaneously form nodules in vitro. Panels a-e depict five distinct human MPM cell lines, SPC111, Meso62, p31, Meso80 and Meso53, after one (a, c, e) or two (b, d) weeks in culture. In comparison, NP3 human primary mesothelial cells remain in a monolayer and do not form aggregates even after 17 days in vitro (f). Asterisks mark nodules, arrow indicates strands interconnecting nodules. 
}
\label{cell_lines}
\end{center}
\end{figure}

\begin{figure}
\begin{center}
\includegraphics[width=5in]{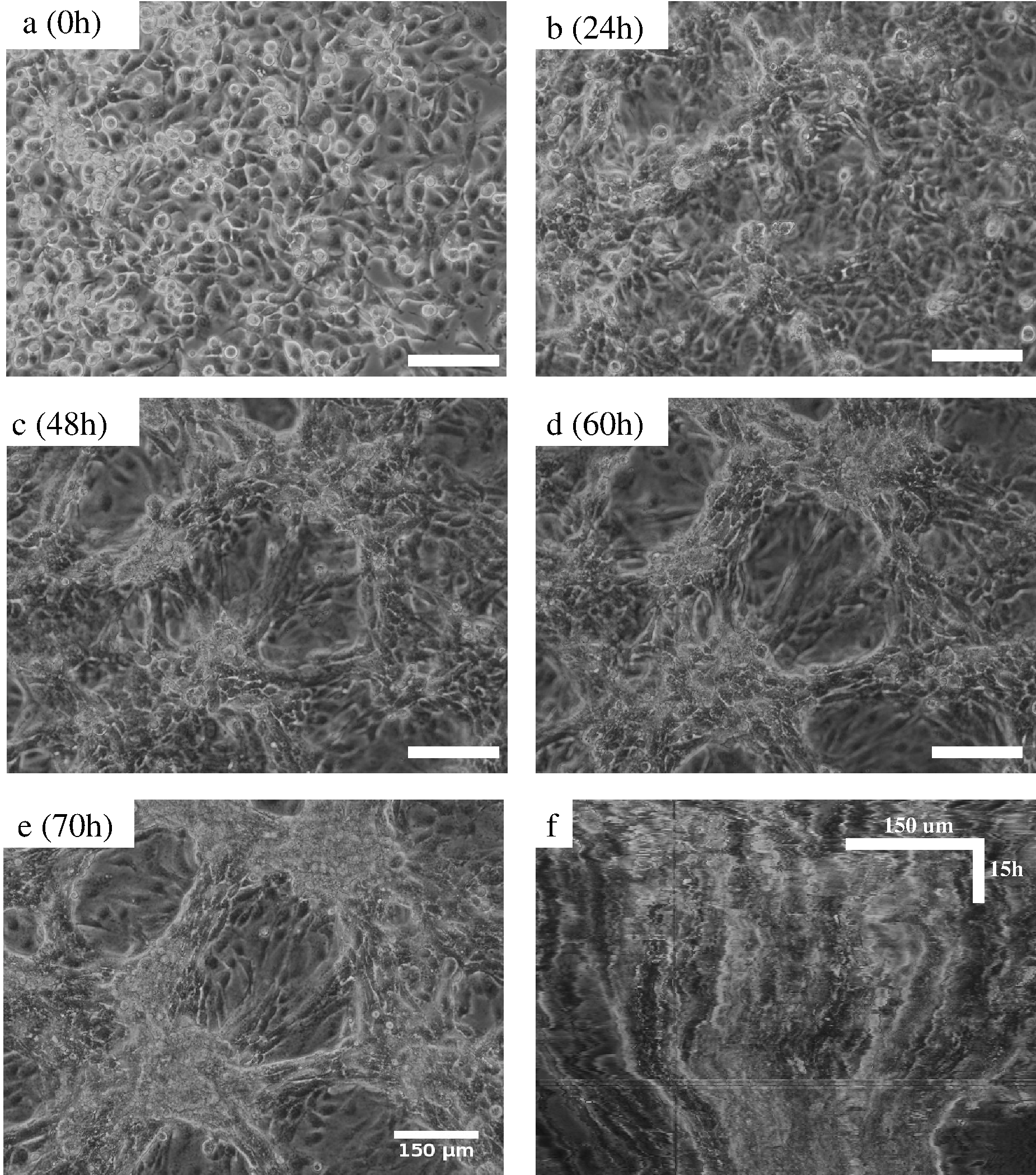}
\caption{\small
Time course of spontaneous nodule formation in a culture of SPC111 cells. Panels a-e are frames from a time-lapse recording taken at seeding, 24h, 48h, 60h and 70h after seeding, respectively. The initial confluent monolayer (a) develops cell density fluctuations (b, c), which eventually become three dimensional nodules interconnected by multicellular strands (d-e). f: The aggregation process is visualized as a kymogram, where pixels along the same line are plotted for each frame. Lines from earlier frames are located at the  top of the image. Scale bars correspond to 150 $\mu$m.
}
\label{development}
\end{center}
\end{figure}

To study the long-term behavior of mesothelioma monolayers, we cultured several, human patient-derived MPM cell lines for up to a few weeks. In this time frame MPM cells form macroscopic multicellular aggregates or nodules, which can be 50 $\mu$m thick and reach a millimeter in lateral extent (Fig.~\ref{cell_lines}). The time needed to generate nodules varies across the cell lines. When SPC111 and p31 cells are seeded at confluency, aggregates reach a macroscopic size after 3-5 days. In contrast, nodule formation in the Meso53, Meso62 and Meso80 lines takes a few weeks. Formation of similar nodules is rather uncommon in cultures of other epithelial cells. Most importantly, under similar culture conditions and duration NP3 human primary mesothelial cells remain in a monolayer (Fig.~\ref{cell_lines}f). Long-term NP3 cultures reach a quiescent state in which both the number of cell divisions and cell deaths are substantially reduced.

Physical cross-sections of the nodules reveal densely packed cells containing prominent, pleiomorphic nuclei. Immunohistochemistry with antibodies against the extracellular matrix protein fibronectin indicates the presence, but not the abundance of ECM in the MPM nodules. Thus, the in vitro MPM nodules are dense, highly cellular structures.

Time-lapse microscopy recordings of the aggregate formation process (Fig.~\ref{development}) reveal that nodules do not form by unusually active cell proliferation. Instead, cells move towards nodules or interconnecting strands. Kymograms (pixels along a selected line plotted for several consecutive frames) indicate a gradual decrease of the extent of an aggregate, a decrease which is approximately linear in in time.

\subsection{Nodule formation is driven by actomyosin contractility}
\begin{figure}
\begin{center}
\includegraphics[width=5in]{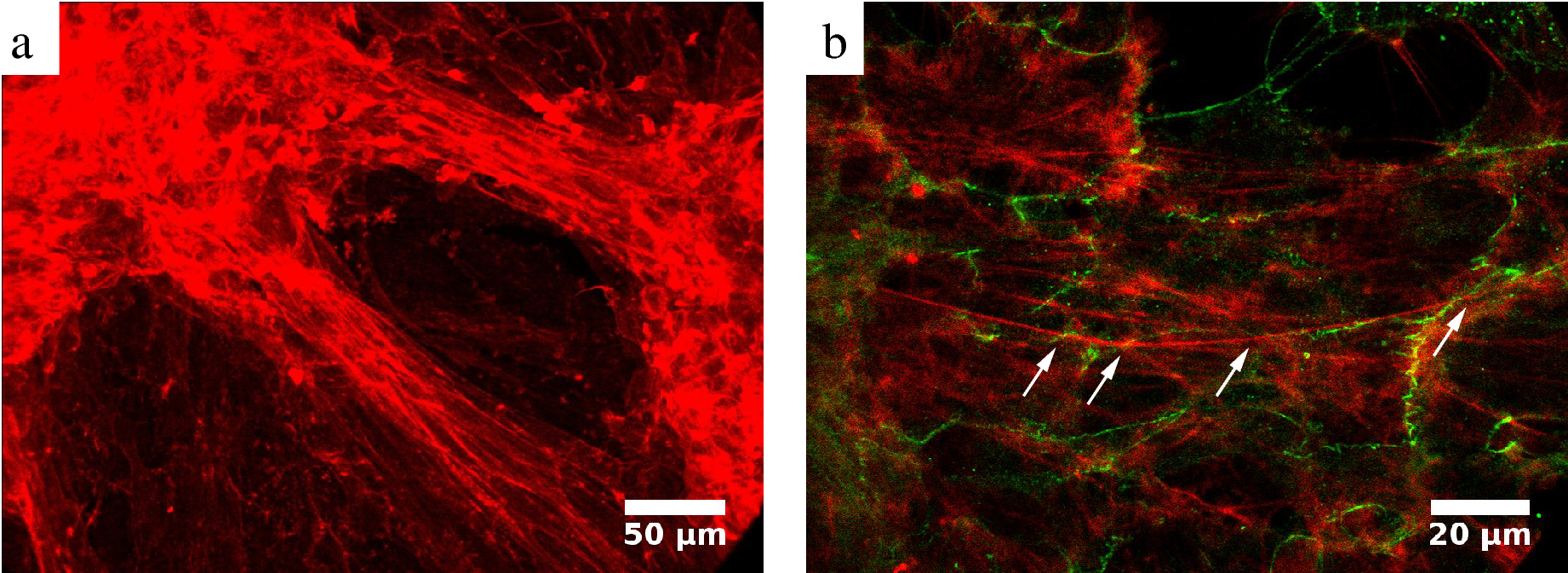}
\caption{\small
In vitro MPM nodules are rich in stress filaments and are mechanically integrated. a: Actin filaments are visualized by confocal microscopy using TRITC-phalloidin (red). Nodules are interconnected with multicellular strands that exhibit parallel bundles of stress filaments. b: Stress cables within multicellular strands align across cell membranes (arrows, visualized by catenin antibodies -- green). 
}
\label{actin_spc111_nodes}
\end{center}
\end{figure}

\begin{figure}
\begin{center}
\includegraphics[width=5in]{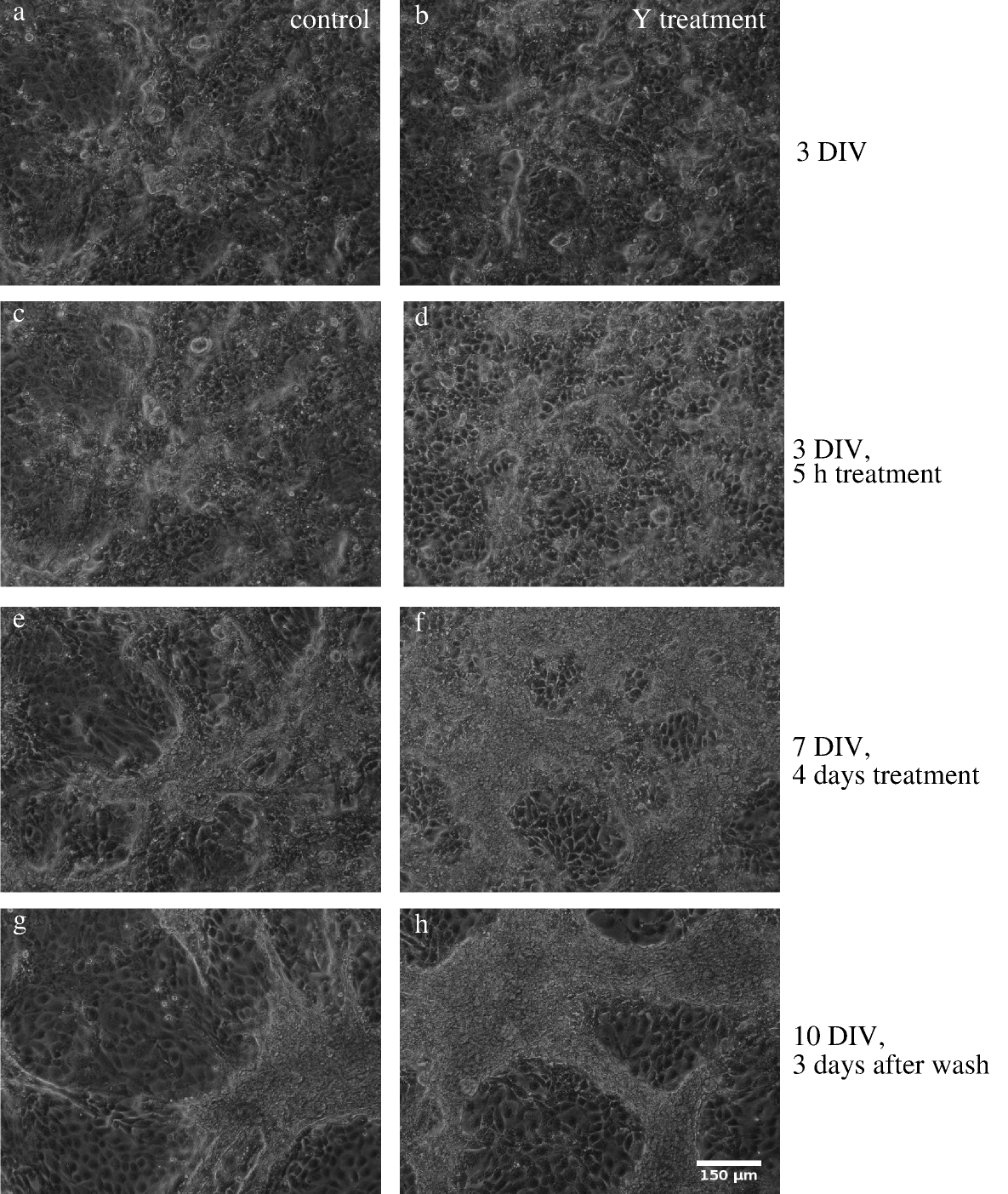}
\caption{\small
In vitro nodule formation requires Myosin II activity. Frames of a time-lapse recording show the morphology of two parallel cultures, an untreated control (a,c,e,g) and one which was transiently exposed to Y27632, a Rho-associated kinase inhibitor (b,d,f,h). The two cultures are similar at 3 days in vitro, at the onset of Y27632 treatment (a,b). Five hours long exposure to Y27632 is sufficient to induce simultaneous flattening and lateral extension of the nodules (d). Cell-dense three dimensional structures, clearly visible in untreated control cultures (e), are largely absent after 4 days of ROCK inhibitor treatment (f). Removal of the drug restores the contractility of MPM cells: three days after replacing the medium nodule morphologies are similar in both cultures (g,h).
}
\label{dev_ROCK}
\end{center}
\end{figure}

Visualization of actin filaments with fluorescent phalloidin (Fig.~\ref{actin_spc111_nodes}) reveals the abundance of stress cables in the strands, organized into parallel bundles connecting the adjacent nodules. Higher resolution confocal images indicate that several stress filaments reach across cell bodies and even form structures that are continuous across cell membranes.

Both the observed cell movements and the presence of profound stress cables within the aggregates suggest that Myosin II dependent cell contractility is an important driving force to collect MPM cells into nodules. To test this hypothesis, we administered drugs that interfere with normal Myosin II activity. Blebbistatin is a potent blocker of acto-myosin contractility as it reduces the affinity of myosin heads to actin \cite{Kovacs04}. The compound Y27632 is a specific inhibitor of Rho-associated kinase (ROCK), one of the activators of Myosin. 

Both blebbistatin (data not shown) and Y27632 (Fig.~\ref{dev_ROCK}) could substantially hinder or completely eliminate both the stress cables and nodule formation in cultures of SPC111 cells. When inhibitors were administered at high concentrations (100 $\mu$M for Y27632 and 80 $\mu$M for blebbistatin), the previously formed nodules flattened: their height decreased by up to 50\% while simultaneously extended laterally. When the inhibitor is present from the onset of the culture, nodule formation can be fully prevented (data not shown). The inhibitors are reversible: aggregation resumes after replacing the medium with fresh DMEM, and three days after medium replacement nodule morphologies are similar in the unperturbed and previously myosin-blocked cultures.

\subsection{ Computational model of contractile cells }
To obtain a computational model of contractile cell sheets, we augmented our elasto-plastic particle model with two new rules: one that controls equilibrium link lengths to maintain a steady intercellular tension, and another to represent adhesions between the cells (particles) and the culture substrate.

We assume that cells strive for a specific contractile environment which they maintain as a homeostatic state: cells increase their contractility if tensile forces between adjacent cells are below a target value, $F^*$. In our model this is achieved by reducing the equilibrium link length $d_l$ in Eq. (\ref{Hook}). Conversely, when tensile forces are too strong, the equilibrium length of the links is increased. In particular, we assume that the rate of change in the link length is proportional to the difference between $F^*$ and $F_l^t$, the tensile component (\ref{fl}) of the force transmitted by the link:
 \be
 {d d_l \over dt} = C d_0 {F_l^t - F^*\over F^*} + \mu_l(t),
 \label{link_adjust}
 \ee
where $1/C\approx 1$ h sets the temporal scale of the feedback regulation. The last term in (\ref{link_adjust}) is an uncorrelated (white) noise, representing random cell shape changes due to factors not considered explicitly in our model. Equation \ref{link_adjust} was integrated over a time interval $\Delta t$, elapsed until the next stochastic event effecting particle connectivity, resulting:
 \be
 d_l(t+\Delta t) = d_l(t) + C d_0 {F_l - F^* \over F^*} \Delta t + \sqrt{D \Delta t}\xi
 \label{noise}
 \ee
where $\xi$ is a pseudo-random variable with unit standard deviation and parameter $D \approx 1 \mu$m$^2/h$ controls the noise amplitude.

The mechanical equilibrium of surface-attached cells requires the net force exerted by adjacent cells to be balanced by the force transmitted through the cell-substrate adhesion complexes to the substrate. Thus, in our model for each particle $i$ we introduce the equilibrium position $\r^0_i$ and angle $\phi^0_i$. The force (and torque) transmitted to the substrate is given by the linear relations
 \be
 \F_{i,0} = k_0 (\r_i - \r^0_i )
 \ee
and
 \be
 \M_{i,0}  = g_0 (\phi_i - \phi^0_i) \e_z.
 \ee
As we discuss below, external forces acting on a cell may contribute to its displacement trough complex and mostly unexplored processes. Here we assume that cells tend to move in the direction of the net external force, hence the eqilibrium position is updated according to
 \be
 {d\r^0_i\over dt} = {\alpha d_0 \over F^0} \F_{i,0} 
 \label{alpha}
 \ee
and
 \be
 {d\phi^0_i \over dt} = {\beta \over d_0 F^0} \M_{i,0}. 
 \ee
Parameters $\alpha$ and $\beta$ set the external force-related bias in cell movements, while the noise in Eq. (\ref{noise}) gives rise to random walk-like movements.

\subsection{ Simulations }

\begin{figure}
\begin{center}
\includegraphics[width=5in]{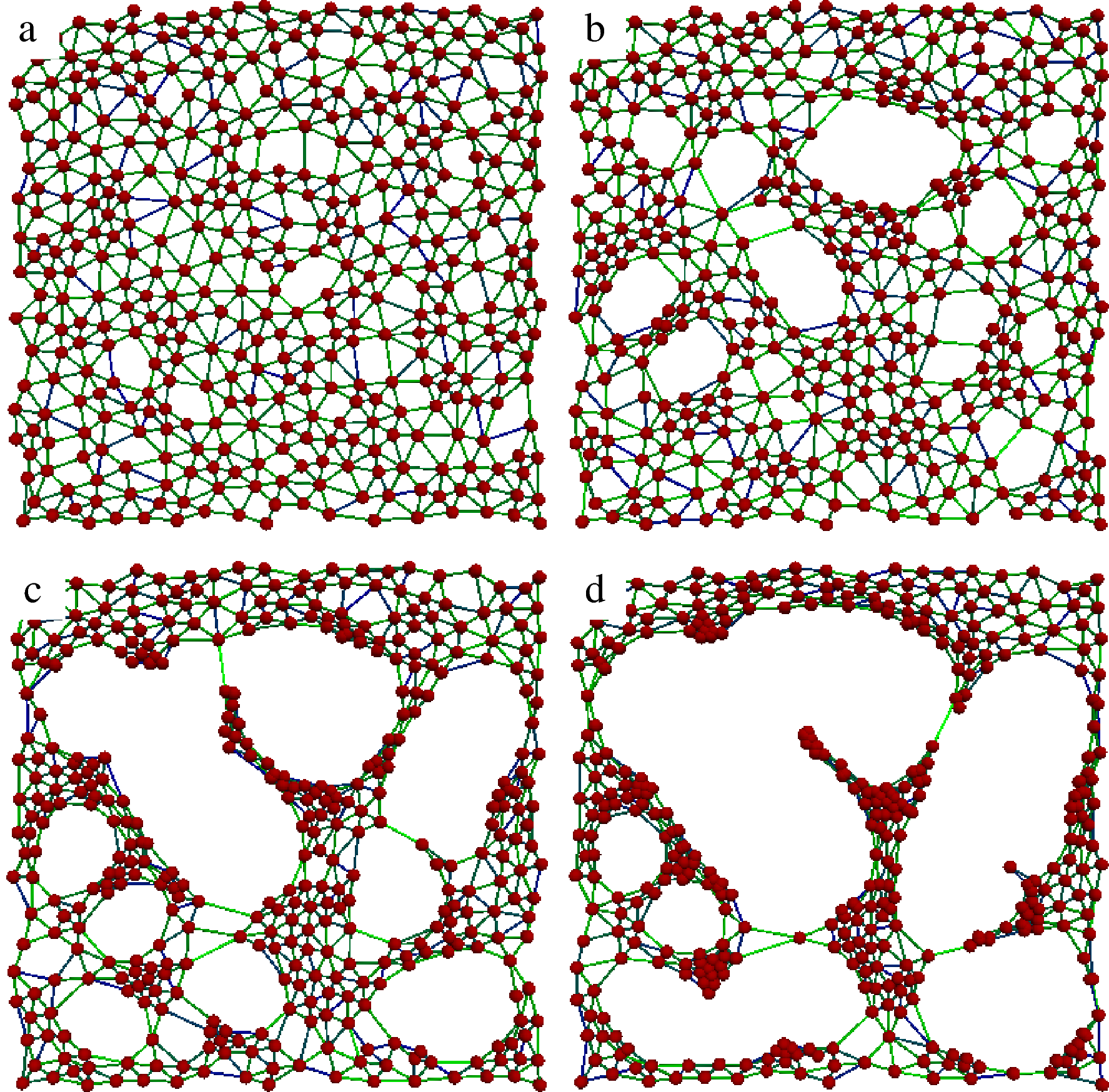}
\caption{\small
Time development of a typical simulation, at 
t=2.5 h (a), t=7.5 h (b), t=13 h (c) and t=16 h (d). Blue to green colors indicate increasing tensile stress within the links. Simulation parameters are listed in table I.
}
\label{model_dev}
\end{center}
\end{figure}

\begin{figure}
\begin{center}
\includegraphics[width=5in]{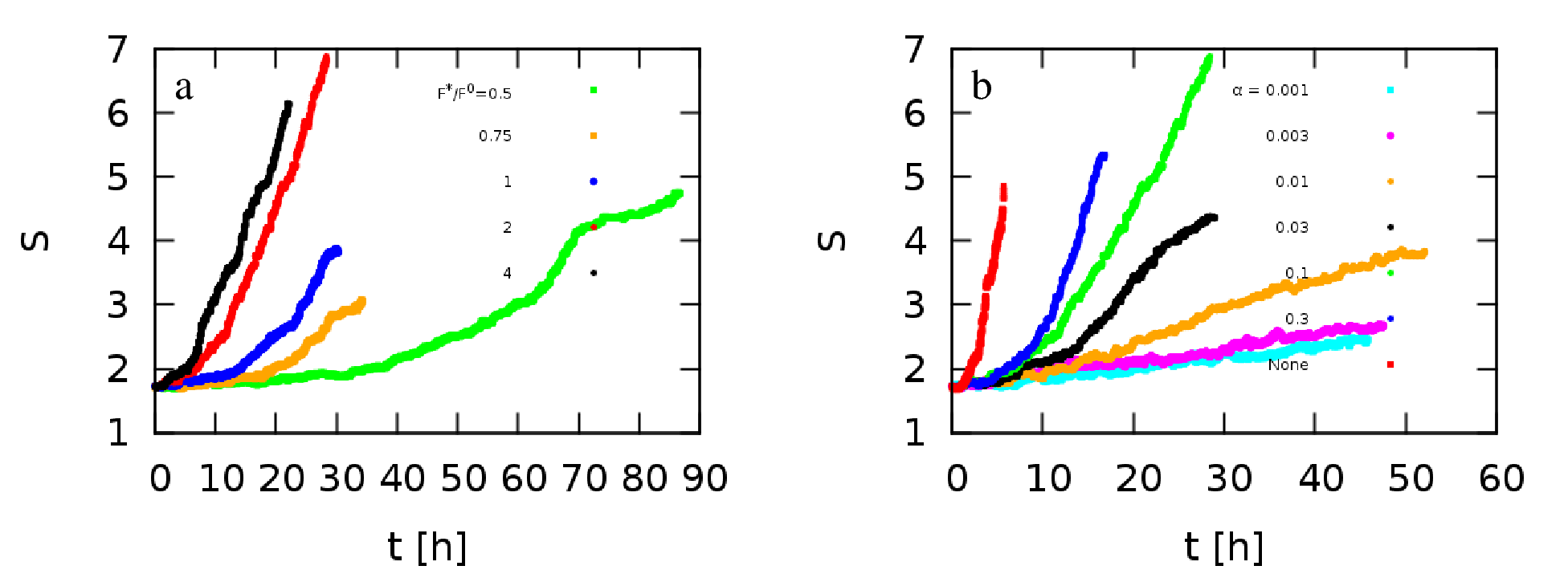}
\caption{\small
The progress of patterning, characterized by the standard deviation of particle density, $S$, as a function of time. Pattern formation is faster for higher target force $F^*$ values (a) and for higher values of $\alpha$, the motion bias towards external forces (b). 
}
\label{param}
\end{center}
\end{figure}

\begin{figure}
\begin{center}
\includegraphics[width=5in]{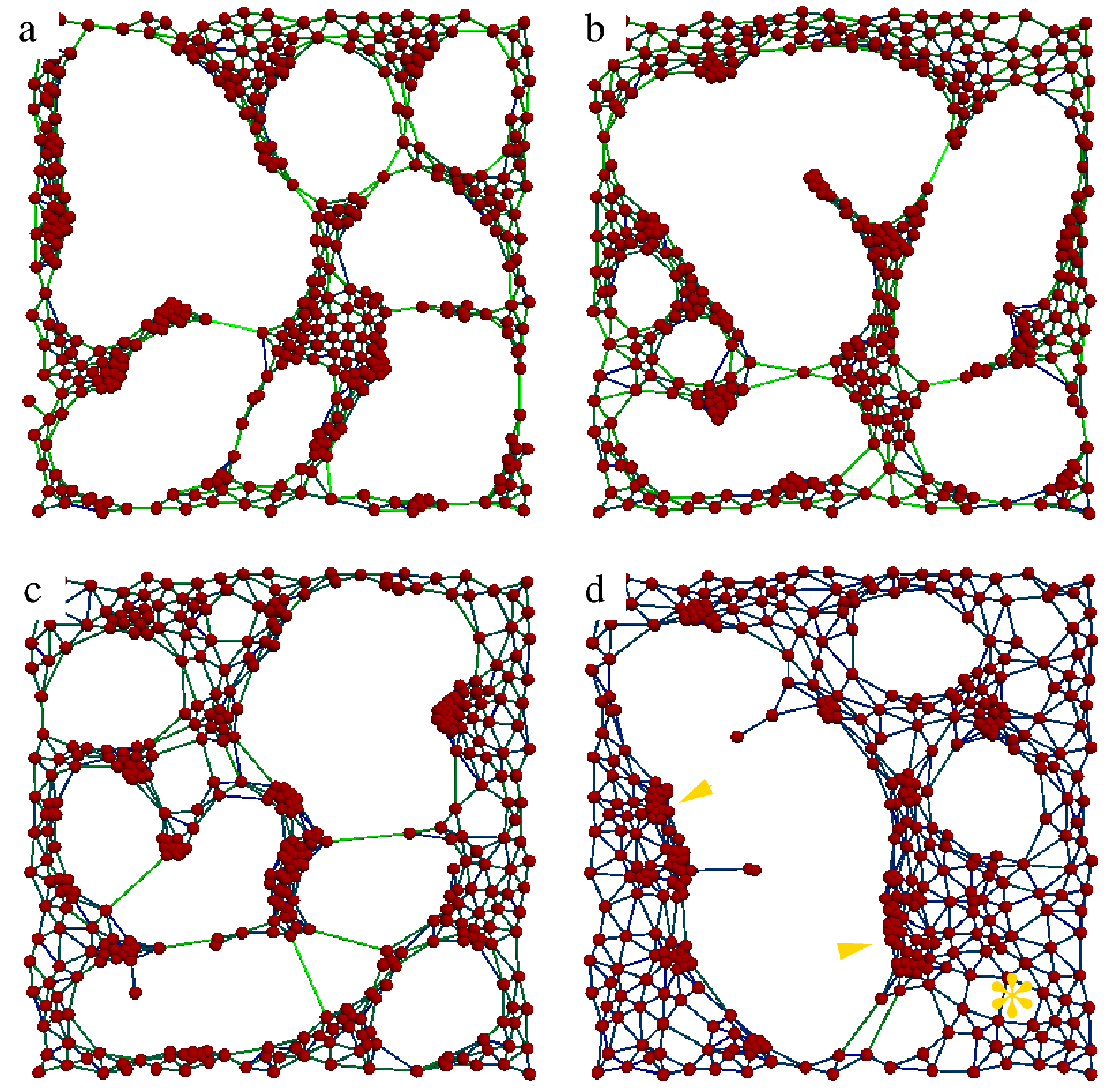}
\caption{\small
Morphologies characteristic for various target force values. Simulations performed with $F^*/F^0=4$ (a), $F^*/F_0=2$ (b), $F^*/F^0=1$ (c), and $F^*/F^0=0.5$ (d) are shown at the same stage of pattern formation ($S=3.85$). For small forces $F^* < F^0$ high cell density clusters develop at the boundary of cell free areas (arrowheads), while the particle density remains low far from such boundaries (asterisk). In contrast, for large forces $F^* > F^0$ the particle density is more uniformly high.
}
\label{Ftgt}
\end{center}
\end{figure}

\begin{figure}
\begin{center}
\includegraphics[width=5in]{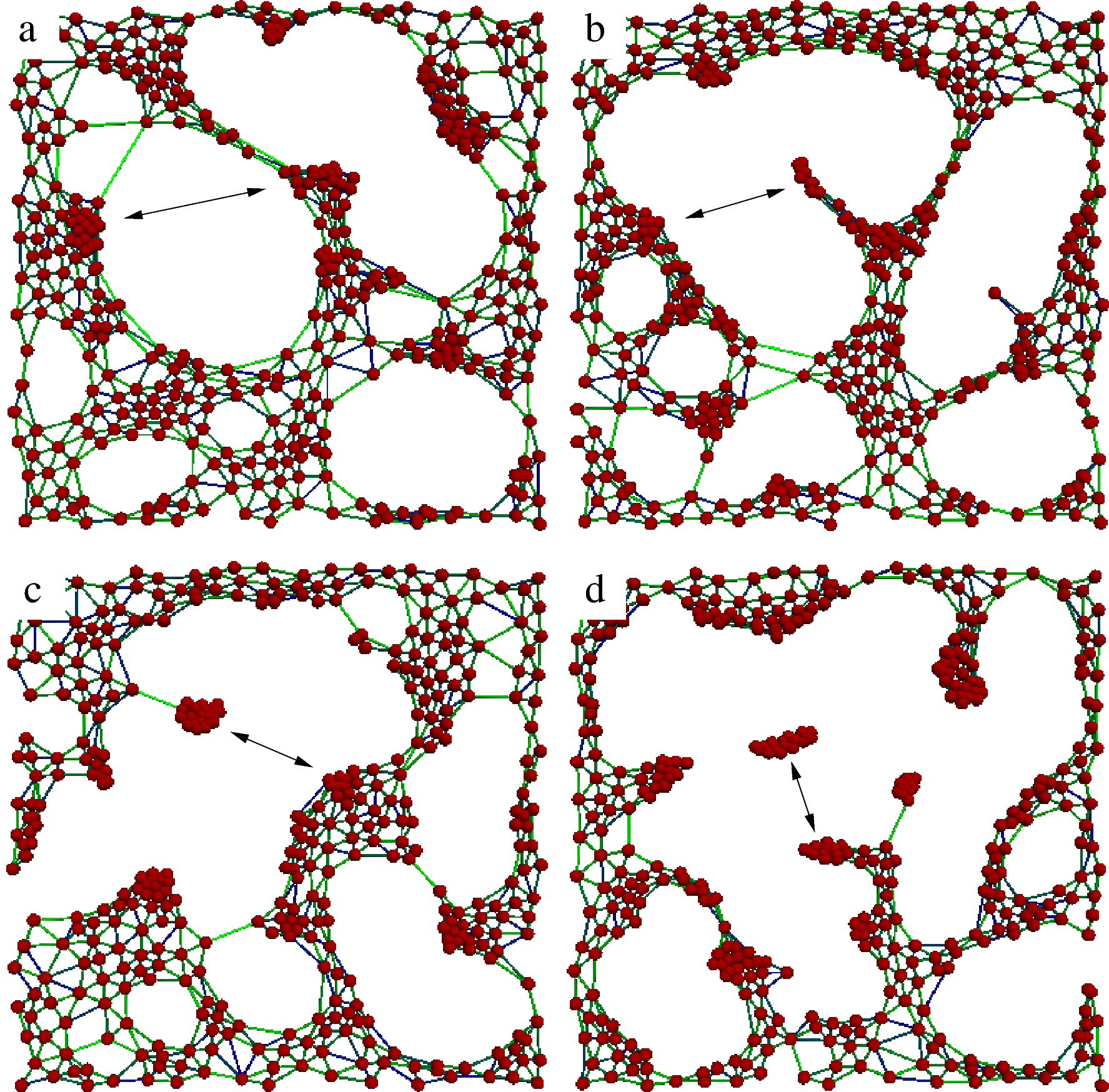}
\caption{\small
Morphologies characteristic for various adhesion parameter values. Simulations performed with $\alpha=0.3$ (a), $\alpha=0.1$ (b), $\alpha=0.03$ (c), and $\alpha=0.01$ (d) are shown at the same stage of pattern formation ($S=3.5$). For smaller values of $\alpha$, the size of the aggregates and the characteristic distance between aggregates (arrows) decreases.
}
\label{adhAdj}
\end{center}
\end{figure}

Simulations of the contractile monolayer (Fig.~\ref{model_dev}) reveal a pattern formation sequence which is in several aspects analogous to the process seen in vitro (Fig.~\ref{development}). First,  the slight initial inhomogeneity of the monolayer is amplified resulting in the formation of holes. Then, cell free areas continue to grow  as their boundary is unstable: links constituting the boundary need to balance the pulling forces exerted by the bulk of the cells. Thus, connections at boundaries experience higher tensile stress which increases both their length and the probability of their removal. Both effects extend the area of the holes. As a consequence, cell density increases in the rest of the system, and eventually cell-covered areas appear as contractile nodes connected by linear strands. While nodules develop in several MPM cultures (Fig.~\ref{development}), the appearance of cell-free areas is observable only in a subset of MPM lines. For example, SPC111 or p31  nodules form on top of a basal monolayer and in such cases our model corresponds to the dynamic upper cell layers. 

During the aggregation process the spatial distribution of cells becomes more inhomogeneous. Thus, the progress of patterning can be characterized by $S$, the standard deviation of the coarse-grained particle density field. While the time course shown in Fig.~\ref{model_dev} is characteristic for all simulations, two model parameters have  important roles in determining both the morphology and the speed of the aggregation (Fig.~\ref{param}). One such parameter is the ratio between the steady state contractile force of the cells, $F^*$, and the Bell threshold of adhesion stability, $F^0$. Figure \ref{Ftgt} compares configurations that are at the same, late stage ($S \approx 4$) of the patterning process. For small forces $(F^* < F^0)$ high cell density clusters develop at the boundary of cell free areas, similar to the in vitro patterns observed for the Meso80 and Meso53 lines (Fig.~\ref{cell_lines}). In this case the aggregation process is slower -- again, in accord with empirical in vitro data. For large forces ($F^* > F^0$) links with high particle density interconnect similarly dense nodules, reminiscent to the patterns observed in SPC111 cultures. 

Model parameter $\alpha$, characterizing the magnitude of external force-directed cell displacements, sets the spatial scale of the pattern (Fig.~\ref{adhAdj}). When cell-substrate adhesions are stable ($\alpha \ll 1$), smaller clusters develop which are close to each other. In contrast, when cells respond strongly to external forces (i.e., substrate adhesion is weak or highly adaptable), fewer and larger clusters form.

\subsection{Linear stability analysis}

To better understand the patterning mechanism we performed a linear stability analysis of the spatially homogenous state. As a continuum model of the multicellular system, we consider a viscoelastic Maxwell material which relaxes shear stresses through an exponential decay, similar to the behavior of the particle model \cite{Czirok14}.  In one dimension the stress distribution $\sigma(x,t)$ is related to the velocity field of the tissue, $v(x,t)$, through the equation
 \begin{equation}
 {\partial \sigma \over \partial t} = \tE {\partial v \over \partial x} - \tC (\sigma - \sigma_*(\rho)) 
 \end{equation}
where $\tE$ is the macroscopic elastic modulus, determined by model parameters $k$ and $g$ \cite{Czirok14}, and $1/\tC$ is the rate of macroscopic stress adjustment analogous to the parameter $C$ in Eq. (\ref{link_adjust}). The first term on the right hand side represents elastic stress and the second term describes the relaxation of the active contractile stress generated by the cells to a value $\sigma_*(\rho)$ that increases with the local cell density as $\sigma_*(\rho) \sim F^*\rho$.

Velocity is set by stress divergence as
 \begin{equation}
 {\partial \sigma \over \partial x}  = \talpha v
 \end{equation}
where the drag coefficient $\talpha$ is analogous to the $\alpha$ pararmeter in Eq. (\ref{alpha}). 

In the absence of cell death and proliferation the cell density $\rho(x,t)$ satisfies the conservation equation
 \begin{equation}
 {\partial \rho \over \partial t} + {\partial \over \partial x} (v\rho) = 0.
 \end{equation}
Considering no-flux boundary conditions the total cell numbers are conserved so that the average cell density $\rho_0$ stays constant in time. Thus the spatially uniform equilibrium solution of the above system is $\rho(x,t) = \rho_0, v(x,t) = 0, \sigma(x,t) = \sigma_0 = \sigma_*(\rho_0)$.

We can analyse the linear stability of the uniform state by considering the time evolution of small perturbations of the form
 \begin{eqnarray}
 \rho(x,t) &=& \rho_0 + \hat\rho e^{st+iqx}\\
 v(x,t) &=& \hat v e^{st+iqx}\\
 \sigma(x,t) &=& \sigma_0 + \hat\sigma e^{st+iqx}
 \end{eqnarray}
where $s$ is the growth rate of a monochromatic perturbation of wavenumber $q$, in the linearised system assuming that the perturbation amplitudes $\hat \rho, \hat v, \hat \sigma$ are small. 

After substitution and neglecting higher then linear terms we obtain
 \begin{eqnarray}
 s\hat \rho + i q \rho_0 \hat v &=& 0\\
 s \hat \sigma &=& i \tE q\hat v - \tC (\hat \sigma - \sigma_*'(\rho_0) \hat \rho)\\
 i q \hat \sigma &=& \talpha \hat \sigma
 \end{eqnarray}
Eliminating the amplitudes we obtain the following quadratic equation for the growth rate $s$
 \begin{equation}
 s^2 + s \left(\tC + \tE q^2{1\over \alpha} \right) -q^2 {\sigma_*'(\rho_0) \tE \tC \over \talpha} = 0
 \end{equation}
with the solutions
 \begin{equation}
 s_{\pm}(q) = {1\over 2} \left ( -b(q) \pm \sqrt{b(q)^2 + 4 q^2 a }\right)
 \end{equation}
where 
 \begin{equation}
 a =  {\sigma_*'(\rho_0) \tE \tC \over  \talpha}
 ,\;\;\; 
 b(q) = \tC + \tE q^2{1\over \alpha}  
 \end{equation}
Since $a$ and $b(q)$ are positive, one of the solutions is always negative while the other is always positive for all wavenumbers. Thus the spatially uniform equilibrium state is unconditionally unstable resulting in increasing cell density fluctuations. This leads to the formation of cell aggregates, however as the perturbations grow the development of the system is not described by the linear approximation since the neglected nonlinearities  become important.

\section{Discussion}
\subsection{Cell displacements and external forces}
As in the case of cell-cell connections, mechanical load acting on cell-substrate adhesion complexes reduces their lifetime \cite{Zhang04}. To relate cell movements and external forces acting on a cell we envision the following process: when a cell-substrate connection breaks, the same mechanical load is distributed along the remaining adhesion complexes. Thus, each of the remaining adhesion sites transmits a larger force, their strain is increased leading to a small displacement of the cell body in the direction of the net external force acting on the cell. Furthermore, when new adhesion complexes form, their equilibrium (stress-free) configuration will correspond to the actual, slightly shifted position of the cell. Thus, by detaching and re-attaching adhesion complexes, the cell relaxes the shear stress between its cytoskeleton and the adhesion substrate, and moves in the direction of the external force. In addition to this purely mechanical connection, external stress may also effect the polarity of active  cell migration \cite{Tambe11,Weber12}.

\subsection{Contraction and aggregation}
Aggregation and sorting involves acto-myosin contractility within the cortex of zebrafish cells \cite{Maitre12}. Here we show that cell groups may also contract through stress cables spanning across several cells, and the resulting system self-organizes into expanding cell-free areas and eventually into free-standing aggregates. Similar behavior also takes place at much smaller length scales in the cytoskeleton of individual cells, where acto-myosin contractility gives rise to f-actin bundles. For example, a similar contractile system, but one that also includes diffusion, has been studied by \cite{Bois11} in the context of pattern formation on the actomyosin cell cortex in which contractility is regulated at molecular level. Further extensions of this problem have been described recently\cite{Kumar14, Moore14}.

\subsection{Future treatment options}
The demonstrated ability of myosin-II inhibitors to flatten mesothelioma nodules may open a new therapeutic method. As cells within avascular, three dimensional nodules are less exposed to systemic drugs, efficient inhibition of nodule formation could enhance the effective drug concentration at the targeted cells. In this case, we would expect a synergistic effect between existing anti-cancer drugs and myosin-II inhibitors.

\section*{Acknowledgements}
This work was supported by the Hungarian Development Agency 
(KTIA AIK 12-1-2012-0041), NIH grant GM102801, and the
and the G. Harold \& Leila Y. Mathers Charitable Foundation.

\section*{References}

\bibliography{cza,cell,ECM,mech,vascular}

\end{document}